\documentclass[%
superscriptaddress,
preprint,
showpacs,
 amsmath,amssymb,
 aps,
floatfix,
]{revtex4}

\usepackage{graphicx}
\usepackage{dcolumn}
\usepackage{bm}

\begin{document}

\title{Prediction of a Stable Post-Post-Perovskite Structure from First Principles}

\author{Changsong Xu}
\affiliation{State Key Laboratory of Low-Dimensional Quantum Physics and Collaborative Innovation Center of Quantum Matter, Department of Physics, Tsinghua University, Beijing 100084, China}%
\affiliation{Physics Department and Institute for Nanoscience and Engineering, University of Arkansas, Fayetteville, Arkansas 72701, USA}%

\author{Bin Xu}
\affiliation{Physics Department and Institute for Nanoscience and Engineering, University of Arkansas, Fayetteville, Arkansas 72701, USA}%

\author{Yurong Yang}
\affiliation{Physics Department and Institute for Nanoscience and Engineering, University of Arkansas, Fayetteville, Arkansas 72701, USA}%

\author{Huafeng Dong}
\affiliation{Department of Geosciences, Center for Materials by Design and Institute for Advanced Computational Science, State University of New York, Stony Brook, NY 11794, USA}%

\author{Artem R. Oganov}
\affiliation{Department of Geosciences, Center for Materials by Design and Institute for Advanced Computational Science, State University of New York, Stony Brook, NY 11794, USA}%
\affiliation{Moscow Institute of Physics and Technology, 9 Institutskiy Lane, Dolgoprudny City, Moscow Region 141700, Russia}
\affiliation{School of Materials Science, Northwestern Polytechnical University, Xi'an 710072, China}

\author{Shanying Wang}
\affiliation{State Key Laboratory of Low-Dimensional Quantum Physics and Collaborative Innovation Center of Quantum Matter, Department of Physics, Tsinghua University, Beijing 100084, China}%

\author{Wenhui Duan}
\affiliation{State Key Laboratory of Low-Dimensional Quantum Physics and Collaborative Innovation Center of Quantum Matter, Department of Physics, Tsinghua University, Beijing 100084, China}%
\affiliation{Institute for Advanced Study, Tsinghua University, Beijing 100084, China}

\author{Binglin Gu}
\affiliation{Institute for Advanced Study, Tsinghua University, Beijing 100084, China}

\author{L. Bellaiche}
\affiliation{Physics Department and Institute for Nanoscience and Engineering, University of Arkansas, Fayetteville, Arkansas 72701, USA}%


\begin{abstract}
 A novel stable crystallographic structure is discovered in a variety of $AB$O$_3$, $AB$F$_3$ and $A_2$O$_3$ compounds (including materials of geological relevance, prototypes of multiferroics, exhibiting strong spin-orbit effects, etc...), via the use of first principles. This  novel structure appears under hydrostatic pressure, and is the first ``post-post-perovskite'' phase to be found. It provides a successful solution to experimental puzzles in important systems, and is characterized by one-dimensional chains linked by group of two via edge-sharing oxygen/fluorine octahedra. Such unprecedented organization automatically results in anisotropic elastic properties and new magnetic arrangements. Depending on the system of choice, this post-post-perovskite structure also possesses electronic band gaps ranging  from zero to $\simeq$ 10 eV being direct or indirect in nature, which emphasizes its ``universality'' and its potential to have striking, e.g., electrical or transport phenomena.
 \end{abstract}


\maketitle

$ABX_3$ perovskites (Pv)  form an important class of crystal structures for which $A$ and $B$ are cations and $X$ is typically the oxygen or fluorine anion. Perovskites display a wealth of phenomena, such as ferroelectricity, magnetism, multiferroicity, piezoelectricity, magneto-electricity, charge and orbital orderings,  superconductivity, etc.... As a result, they constitute a rich playground for research and are important for various technologies, which explains the flurry of activities that have been devoted to them \cite{mri.4.3}. Interestingly, recent works have shown that applying a hydrostatic pressure in some $ABX_3$ materials can result in the transformation from the Pv structure to the so-called ``post-perovskite'' (pPv) structure which can have important physical consequences \cite{nature.430.445,jssc.4.275,pepi.165.127,pcm.36.455,martin,martin2,pcm.35.189,science.304.855}. For instance, the pPv structure discovered in MgSiO$_3$ explains the existence of  anisotropic features  in the  D" layer of Earth \cite{nature.430.445,nature.438.1142,science.304.855}.
 Moreover, CaRhO$_3$ was recently found to adopt a polymorph that was described as being an intermediate phase between perovskite and post-perovskite \cite{Shikaro}.
Based on these discoveries, one may wonder if  there is another crystal structure (to be termed as ``post-post-perovkite'' (ppPv)) for which Pv or pPv materials can evolve to under hydrostatic pressure. Positively answering such question will deepen the current knowledge of crystallography and high pressure. Moreover, if such structure does exist, one may also wonder about its structural characteristics and if they can lead to novel physical properties -- which  is obviously interesting for fundamental reason but also for the design of original devices. It is also of high importance to determine what precise compounds may possess such hypothetical structure. In particular, could it be that such structure is not only the answer to some experimental puzzles
 in some compounds that are of high geological relevance  \cite{martin,martin2} but can also exist in a variety of materials (ranging from metals to insulators, via multiferroics and systems possessing strong-spin-orbit effects or rare-earth ions)? If that is the case, this structure is ``universal'' and has the high potential to lead to the discovery of many striking phenomena.

The goal of this Letter is to address all these aforementioned unknown questions, via the use of
first-principles calculations. As we will see, surprises are in store since we, e.g.,  (1) predict that many and various {\it $ABX_3$} and $A_2$O$_3$ materials can transform to a common, novel and stable ppPv structure under hydrostatic pressure; and (2)  reveal its unusual structural, magnetic and electronic properties. Moreover, this ppPv structure is likely the ``mysterious'' phase that has been observed in Refs. \cite{martin,martin2}.

 As detailed in the Supplementary Materials (SM), first-principles calculations are performed on many {\it $ABX_3$} and $A_2$O$_3$ materials, with different $A$ and $B$ atoms and with $X$ = O or F anion, under hydrostatic pressure. A list of these materials is indicated in Fig. 1.

 {\it Crystal structures.} Let us first concentrate on a specific material that has been experimentally explored under pressure, namely NaMgF$_3$.
Figure 2(a) shows that the orthorhombic Pv $Pnma$ phase (Pv-$Pnma$) is predicted to be its ground state up to $\simeq$ 20 GPa, as consistent with measurements \cite{pcm.4.299,martin,martin2}.
Such phase is common to many perovskites \cite{acta.sryst.60.263} and is schematized in Fig. 2(b).
In this phase, any fluorine (or oxygen) octahedra  share corners with their neighboring octahedra along the pseudo-cubic [100], [010] and [001] directions.
Figure 2(a) further reveals that NaMgF$_3$ is predicted to experience a phase transition to the  (orthorhombic) {\it post-perovskite} $Cmcm$ phase (pPv-$Cmcm$) at $\simeq$ 20 GPa, for which not only the space group but also the crystallographic structure change, as schematized in Fig. 2(c).
Interestingly, the Pv-$Pnma$--to--pPv-$Cmcm$ transition has been observed to occur for pressure around 27-30 GPa and under laser heating (likely, to overcome the kinetic barrier inherent to first-order transitions) \cite{martin,martin2} in NaMgF$_3$, which is rather consistent with our prediction of a corresponding critical pressure of $\simeq$ 20 GPa at 0 Kelvin. As indicated by Fig. 2 and Table I of the SM,   the pPv-$Cmcm$ phase differs from the  Pv-$Pnma$ structure by the existence of  two-dimensional sheets formed by  octahedra that share edges
along the $a$-axis and  corners along the $c$-axis. These two-dimensional sheets are stacked
along the $b$-axis with an interlayer made of $A$ atoms separating any two neighboring sheets.
As a result, the elastic (stiffness) constant of  pPv-$Cmcm$ is much lower along the $b$-axis
than along the $a$ or $c$ axis for any material,  including NaMgF$_3$ (see Table II of the SM)  and MgSiO$_3$ -- which, for this latter compound, is consistent with the seismic anisotropy observed in the so-called D" layer of Earth \cite{nature.430.445,nature.438.1142}.

As also revealed by Fig. 2(a), we further found that NaMgF$_3$ undergoes another  transition at $\simeq$ 51 GPa, for which the space group and crystallographic structure both change again:
the resulting phase re-adopts the $Pnma$ space group but within a different crystallographic structure that is  termed  ``post-post-perovskite''  \cite{prb.76.184101,ic.51.6559,martin2} and that is
denoted as ppPv-$Pnma$ in the following. Its structural characteristics are shown in Figs. 2(d) and 2(e). Interestingly, while ppPv-$Pnma$ has never been previously reported in any material, its present discovery solves a puzzle: it likely is the so-called  mysterious ``N-phase''   that has been observed in Ref.\cite{martin,martin2}, based on the facts that (i) it experimentally appears as a result of a phase transformation from the pPv-$Cmcm$ phase at 56 GPa under  laser-heating of about 2000 K (as consistent with our  predicted pPv-$Cmcm$--to--ppPv-$Pnma$ transition  for a critical pressure $\simeq$ 51 GPa at T = 0 K);  (ii) the ``N-phase'' has been assigned an orthorhombic symmetry \cite{martin}, in line with the $Pnma$ space group we presently found for our ppPv structure \cite{footnotesymmetry}; and (iii)  our simulated X-Ray Diffraction pattern of ppPv-$Pnma$ is consistent with the one experimentally found in Ref.\cite{martin2} for this N-phase (see Fig. 3 of the SM).

Remarkably, comparing Figs. 2(c) with 2(d) and 2(e) reveals that the transformation from
pPv-$Cmcm$ to ppPv-$Pnma$ results in the breaking of the two-dimensional octahedra sheet at the shared corners in favor of  one-dimensional chains that are elongated along the $b$-axis of the ppPv-$Pnma$ structure.
These chains organize themselves by group of two (with the two chains forming the double chain being parallel to each other along the $b$-axis), as a result of edge-sharing octahedra.
As shown in Table I of the SM, for a given pressure of 60 GPa (which is rather close to the predicted pPv-$Cmcm$--to--ppPv-$Pnma$ transition),
the formation of these  double chains leads, in NaMgF$_3$, to the $b$ and $c$ lattice constants of ppPv-$Pnma$
increasing by 4.3\% and 25.3\%, respectively, with respect to the $a$ and $b$ lattice constants  of pPv-$Cmcm$. On the other hand,  the $a$ lattice parameter of  ppPv-$Pnma$ decreases by 24.9\% with respect to the $c$ lattice constant of pPv-$Cmcm$  (note that the $b$-axis is parallel to the chains in ppPv-$Pnma$ while it is  perpendicular to the octahedra sheets in pPv-$Cmcm$, implying that comparisons have to be made between the ($a,b,c$) triad axis of ppPv-$Pnma$ and the ($c,a,b$) triad axis of pPv-$Cmcm$).
Such changes in lattice constants  result in a decrease of 1.84\% of the volume at the pPv-$Cmcm$--to--ppPv-$Pnma$ transition in NaMgF$_3$,
which is a prediction that can be easily checked by measurements. Note also that the octahedra  are more distorted in  ppPv-$Pnma$  than in pPv-$Cmcm$, as evidenced by the facts that the six Mg-F bonds of the octahedra in ppPv-$Pnma$ adopt four different values equal to  1.813 \AA, 1.871 \AA ~(doubly degenerate),  1.888 \AA ~(doubly degenerate) and 1.942 \AA, respectively, while those of pPv-$Cmcm$ only split between two values of 1.785 \AA ~ (doubly degenerate) and 1.846 \AA ~(four times degenerate), respectively, for a pressure of 60 GPa. The fluorine octahedra therefore become 0.84\% larger
in ppPv-$Pnma$  than in pPv-$Cmcm$ (even if the volume decreases), as edge-sharing allows for more compact packing.
Moreover, in the ppPv-$Pnma$ phase,  any Mg ion belonging to one chain gets rather close to a specific F ion belonging to the adjacent chain (indicated by the dashed line in Fig. 2(d)) forming the double chains and therefore leads to an increase in coordination number from 6 to ``6+1''. For instance, at 60 GPa, the bond between these Mg and specific F ions is about 2.103 \AA, which is comparable to the distances of 1.813 \AA - 1.942 \AA~ between Mg and F ions belonging to the same octahedra \cite{footnorecomparison}.

Figure 1(a) also shows that many materials are also predicted to exhibit the aforementioned Pv-$Pnma$--to--pPv-$Cmcm$ and pPv-$Cmcm$--to--ppPv-$Pnma$ transitions,  but at different critical pressures. On the other hand, Fig. 1(a) further indicates that some materials are predicted to {\it directly} transform from Pv-$Pnma$ to ppPv-$Pnma$ without adopting the intermediate pPv-$Cmcm$ phase, as the pressure increases. Examples include (i) two prototypes of multiferroic materials, BiFeO$_3$ and BiCrO$_3$ \cite{sicence.299.1719,apl.88.152902}; (ii) CaMnO$_3$ that has been predicted to exhibit both magnetic and electric orderings when grown as a strained film  \cite{prl102.117602,prl.107.197603}; and (iii) the rare-earth ferrites $R$FeO$_3$ \cite{White1969,Hongjian2013,prb.89.205122} with small or intermediate ionic radius. For instance, GdFeO$_3$ directly undergoes a transition from Pv-$Pnma$ to ppPv-$Pnma$ at the pressure of $\simeq$ 56.5 GPa.
Conversely, there are some materials, such as Ca$B$O$_3$ with $B$ = Ru, Ir, Rh, Pt
(that have been investigated because of their analogy with MgSiO$_3$ \cite{jssc.4.275,grl.32.l13313,pepi.165.127,pcm.36.455,pcm.35.189,ic.47.1868} or because of the strong effect of spin orbit interactions on some of their physical properties \cite{prl.110.217212}) that do not exhibit the Pv-$Pnma$ phase but rather evolve from pPv-$Cmcm$ to ppPv-$Pnma$, as a hydrostatic pressure is applied and increased. In particular, we predict that the ppPv-$Pnma$ phase of CaRuO$_3$ will appear at a pressure of 33.8 GPa, which should make its observation rather easily feasible.
On the other hand and as  shown in Fig. 1(b), no ppPv structure was found {\it up to 120 GPa} in some other systems, such as $R$FeO$_3$ compounds with large ionic radius (i.e., $R$ = Nd, Pr, Ce and La),  MgSiO$_3$, Mn$_2$O$_3$ or Al$_2$O$_3$ -- as consistent with measurements and
previous computations \cite{science.304.855,nature.430.445,Santillan,epsl.246.326,gpl.32.l16310} (note that the SM provides a more detailed comparison between our predictions and these previous works).

 {\it Dynamical stability.} The ppPv-$Pnma$ structure is {\it dynamically} stable in its pressure range of stability for all the materials shown in Fig. 1(a). Two examples are shown in Figs. 1(a) and 1(c) of the SM for NaMgF$_3$ and GdFeO$_3$, respectively,  both under a pressure of 60 GPa. In fact, we also numerically found that, in several studied compounds, ppPv-$Pnma$ does not have any unstable phonon even  in pressure regions for which this phase is not the lowest one in enthalpy.
  For instance, ppPv-$Pnma$  is dynamically stable even at zero pressure in, e.g., CaPtO$_3$, which likely implies that this phase can be  quenched to ambient pressure in this material (especially because the difference in enthalpy between  pPv-$Cmcm$ and ppPv-$Pnma$ is found to be as small as 181 meV/5-atom at zero pressure in CaPtO$_3$).
Conversely, other phases, such as pPv-$Cmcm$, can also have no unstable phonon in the
pressure range for which ppPv-$Pnma$ has the lowest enthalpy, which implies that (i) pPv-$Cmcm$ may still be experimentally found in some materials at pressure  higher than the predicted pPv-$Cmcm$--to--ppPv-$Pnma$ transition pressure and (ii) observing ppPv-$Pnma$ phase in these materials may require the use of laser heating (to overcome kinetic barrier).

{\it Electronic structure.} We also numerically found that, within ppPv-$Pnma$,  the {\it electronic} band gap can be rather quantitatively different between investigated materials (see Table III and Fig. 1 of the SM). For instance,  the calculated band gap of NaMgF$_3$  is as large as 9.04 eV for a pressure of 60 GPa while it is dramatically reduced to 0.83 eV for GdFeO$_3$ under the same pressure.  In fact, a few systems are even metallic above the pressure at which the ppPv-$Pnma$ phase begins to appear. Examples include CaRhO$_3$ at 70 GPa and  CaIrO$_3$ at 90 GPa.
Equally striking and as shown in Fig. 1 of the SM too, even the character of the band gap (that is direct {\it versus} indirect) can be altered when going from one material to another within  ppPv-$Pnma$. Such electronic flexibility may result, in the future, to the discovery of anomalous properties (such as metal-insulator transitions \cite{rmp.70.1039}) or highly-desired features (such as a direct-band gap in the frequency spectrum needed for photovoltaic devices \cite{future.34.663} or light-emitting devices \cite{nature.347.539}) in materials possessing the  ppPv-$Pnma$ structure.

{\it Magnetic ordering.}  Interestingly, some $AB$O$_3$ materials, that are predicted to exhibit   ppPv-$Pnma$ structure, possess $A$ and/or $B$ atoms that are {\it magnetic}. As a result, novel or striking magnetic arrangements should emerge, especially when recalling that  ppPv-$Pnma$ adopts unusual ``double'' one-dimensional chains inside which $A$ and $B$ bond with O atoms (see Figs. 2(d) and 2(e)).  Let us, for instance, consider the case of the ppPv-$Pnma$ phase of GdFeO$_3$ at 60 GPa and include the 4$f$ electrons of Gd in the valence in the calculations, thus allowing both Gd and Fe ions to adopt localized magnetic moments (that are found to be 6.90 $\mu_B$ and 4.12 $\mu_B$, respectively).
Practically,  enthalpies of different collinear magnetic configurations are computed and used to extract the coupling coefficients ($J_{BB,chain}$,  $J_{BB,across}$, $J_{BA,single}$, $J_{BA,four}$) of  the  model described by $H$ = $\frac{1}{2}J_{BB,chain}$$\sum$$_{i,j}${$\bm{S}_i$$\cdot$$\bm{S}_j$} + $\frac{1}{2}J_{BB,across}$$\sum$$_{i,j}${$\bm{S}_i$$\cdot$$\bm{S}_j$} + $\frac{1}{2}J_{BA,single}$$\sum$$_{i,j}${$\bm{S}_i$$\cdot$$\bm{S}_j$} + $\frac{1}{2}J_{BA,four}$$\sum$$_{i,j}${$\bm{S}_i$$\cdot$$\bm{S}_j$}, where the sums over $i$ run over all Fe atoms while the first (respectively, last) two sums over $j$ run over specific Fe (respectively Gd) atoms that will be indicated below.
As depicted in Fig. 3,  the strongest coupling coefficient (denoted by $J_{BB,chain}$) is found to be 2.86 meV (that is antiferromagnetic in nature) and is between Fe ions that are distant (by $\simeq$ 3.02 \AA) along the $b$-axis. Interestingly, the coupling between
Fe ions that belong to two adjacent and parallel one-dimensional chains (and are distant by $\simeq$ 2.67 \AA) is also antiferromagnetic in nature but is of smaller magnitude since it is equal to
 1.52 meV (this parameter is denoted here as $J_{BB,across}$). As a result and as shown in Fig. 3(a), the magnetic ground state of GdFeO$_3$ possesses one-dimensional antiferromagnetic chains elongating along the $b$-axis and formed by Fe ions  with each of these Fe ions having two neighboring Fe ions of opposite spins and that belong to the adjacent parallel chain.
 Note that the particular {\it triangular-like} geometry seen by any  magnetic $B$ ion (see Fig. 3(a)) because of the formation of the double one-dimensional chains inherent to ppPv-$Pnma$ in $AB$O$_3$ materials is a perfect ``recipe'' to generate the so-called {\it geometric frustration} \cite{PhysToday2006,nature.470.513} in the specific (and presently hypothetical) case that $J_{BB,chain}$ and $J_{BB,across}$ would still be antiferromagnetic in nature but would now be close to each other in magnitude (unlike in GdFeO$_3$). Searching for such compounds or  the hypothetical pressure giving rise to such condition in some materials therefore constitutes a promising avenue to pursue in the future.
 Note also that we numerically found that, in  the ppPv-$Pnma$ phase of GdFeO$_3$ at 60 GPa,
 magnetic interactions between Gd ions are negligible (as consistent with the deep $f$-shell of Gadolinium) but Fe ions are antiferromagnetically coupled with their closest Gd ions.
 As indicated in Fig. 3(b), the resulting coupling is $J_{BA,single}$ = 1.22 meV between Fe and Gd ions that form single bond (and are distant by 2.783 \AA) while it is $J_{BA,four}$ = 0.73 meV
 between Fe and Gd ions that are tetrahedrally bonded (and distant by 3.121 \AA).
 As a result, the magnetic ordering of Gd ions is governed by their interaction with Fe ions and is the one depicted in Fig. 3(b).

 In summary, we used first-principles techniques to discover a common and stable ppPv crystal structure in a variety of $ABX_3$ and $A_2$O$_3$ materials under pressure. Such phase exhibits one-dimensional structural characteristics which naturally lead to strong anisotropy and emergence of novel magnetic orderings, and provides a plausible explanation for the mysterious  phase  that has been reported in Refs. \cite{martin,martin2}.  Moreover, the electronic band gap of this phase is highly dependent on the system and can be of  rather different nature and magnitude, which points towards the ``universal'' aspect of this new phase.  We  hope that  this Letter will encourage researchers to confirm the predictions presently reported and to determine  properties associated with such novel crystal structure.

This work is financially supported by the Department of Energy, Office of Basic Energy Sciences, under contract ER-46612  and ONR Grants N00014-11-1-0384 and N00014-12-1-1034. It is also supported by the Ministry of Science and Technology of China (Grant No. 2011CB606405) and National Natural Science Foundation of China (Grant No. 11174173). We also acknowledge ARO Grant W911NF-12-1-0085  and  NSF grant  DMR-1066158  for discussions with scientists sponsored by these grants. The calculations were performed on the ``Razor" (Univ. of Arkansas) and ``Explorer 100" (Tsinghua Univ.) cluster systems.


\clearpage
\begin{figure*}
\includegraphics[width=16cm]{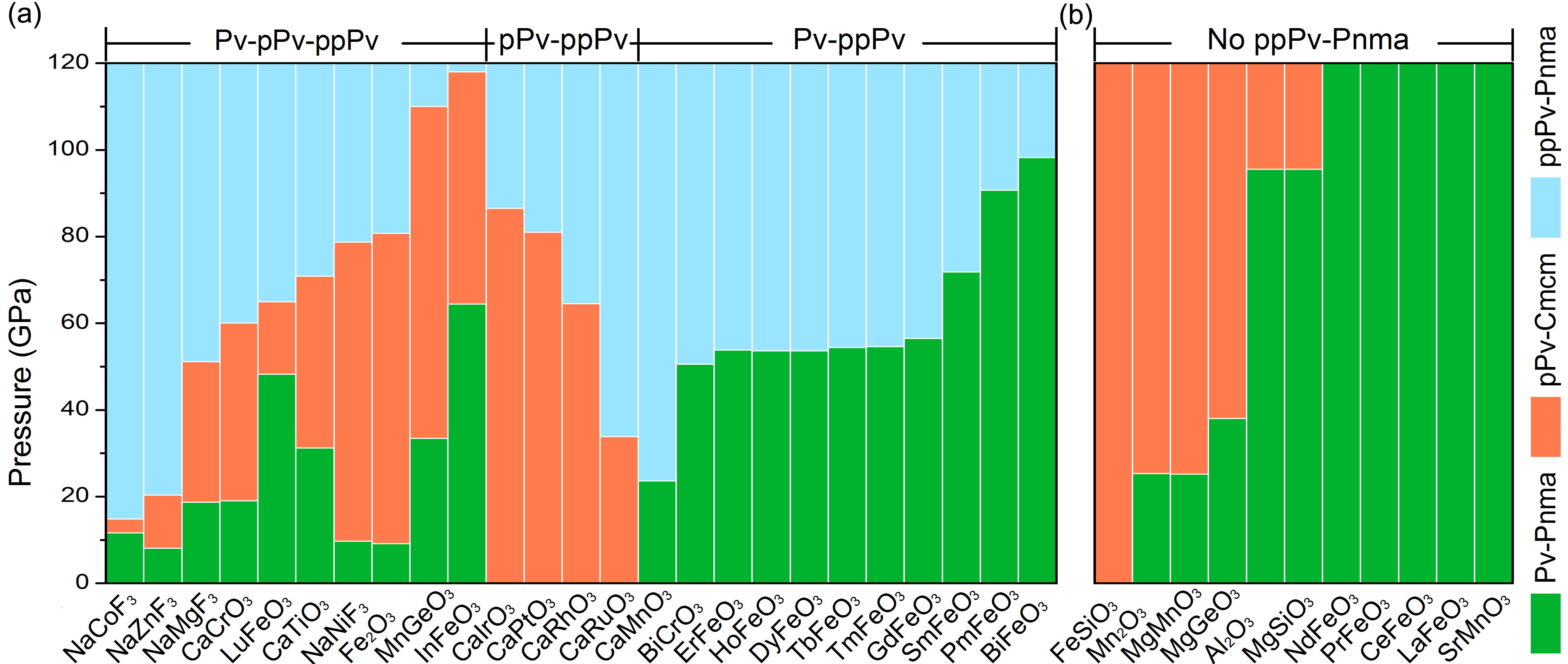}%
\caption{(Color online). Pressure range of stability of the Pv-$Pnma$, pPv-$Cmcm$ and ppPv-$Pnma$ phases in the
{\it ABX$_3$} and $A_2$O$_3$  materials under study. Panels (a) and (b) report materials possessing or missing, respectively, the  presently discovered ppPv-$Pnma$ structure for pressure up to 120 GPa.}
\end{figure*}

\begin{figure*}
\includegraphics[width=16cm]{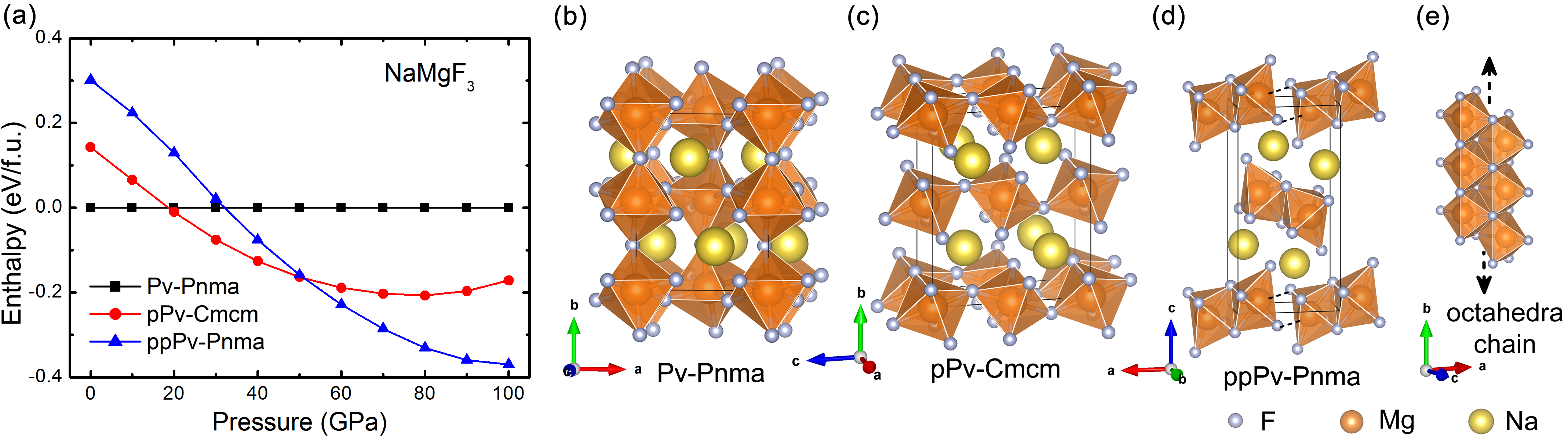}%
\caption{(Color online). Pressure dependence of the enthalpy of the Pv-$Pnma$, pPv-$Cmcm$ and ppPv-$Pnma$ phases of NaMgF$_3$ (Panel (a)), along with the schematization of (b) the Pv-$Pnma$, (c) pPv-$Cmcm$ and (c) and (d) ppPv-$Pnma$ crystallographic structures. Note that the enthalpy of the Pv-$Pnma$ phase has been set to be zero for any pressure in Panel (a).}
\end{figure*}

\begin{figure}
\includegraphics[width=8cm]{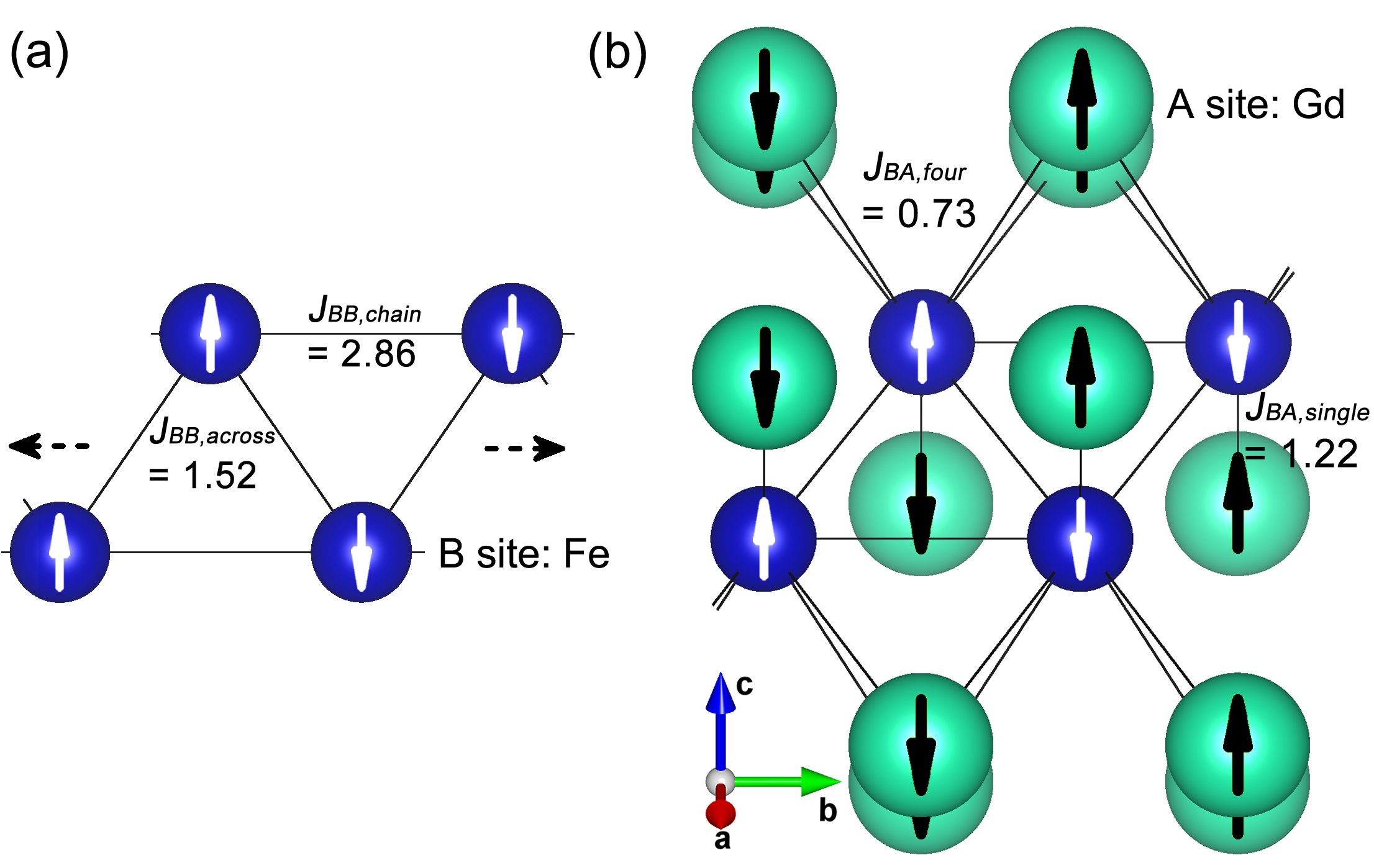} %
\caption{(Color online). Magnetic ground state of the ppPv-$Pnma$ phase of GdFeO$_3$ under 60 GPa.
Panel (a) reports the strength of the magnetic interactions and the resulting magnetic ordering between Fe ions, while Panel (b) depicts the coupling coefficients  associated with Fe and Gd magnetic interactions as well as the spin pattern adopted by these two types of ions.}
\end{figure}


\begin{references}
\bibitem{footnotesymmetry} Note, however, that Ref. \cite{martin} tentatively  assigns a $Pnnm$ space group rather than $Pnma$ for their N-phase.  It will thus be interesting if the authors of Ref. \cite{martin}  can check if $Pnma$ better describes their experimental data, especially once realizing that $Pnnm$ and $Pnma$ share the same $mmm$ point group.
\bibitem{footnorecomparison} Note that,  for comparison, the closest distance between Mg and F ions that do not belong to the same octahedra is about  2.929 \AA~ in the pPv-$Cmcm$ phase at around 60 GPa, while the distances between Mg and F ions belonging to the same octahedra range between 1.785 \AA~  and 1.846 \AA.
\end{references}
\end{document}


\title{Supplemental Material for \\  ``Prediction of a Stable Post-Post-Perovskite Structure from First Principles''}

\author{Changsong Xu}
\author{Bin Xu}
\author{Yurong Yang}
\author{Huafeng Dong}
\author{A. R. Oganov}
\author{Shanying Wang}
\author{Wenhui Duan}
\author{Binglin Gu}
\author{L. Bellaiche}

\maketitle

The goal of this Supplementary material (SM) is to provide details about the methods used, as well as additional information.

{\it Methods:}

Chronologically, this study started by applying the {\it ab-initio} evolutionary algorithm USPEX  \cite{jcp.124.244704,cpc.184.1172,acr.44.227,rmg.71.271} to LuFeO$_3$. Calculations of (LuFeO$_3$)$_n$ with 10, 20 and 30-atom cells (corresponding to n = 2, 4 and 6, respectively) were performed at 60 GPa and 100 GPa. We successfully reproduced the occurrence of the Pv-$Pnma$ phase of LuFeO$_3$, which is often referred to as the GdFeO$_3$-type structure and is known to be present in many perovskites \cite{acta.sryst.60.263}. We further predicted the existence of two other stable crystallographic structures of low enthalpy, viz. pPv-$Cmcm$ and ppPv-$Pnma$, in LuFeO$_3$. Then, the various {\it ABX$_3$} and $A_2$O$_3$ materials indicated in Fig. 1 of the manuscript were systematically studied under hydrostatic pressure, using the Vienna {\it ab-initio} simulation package (VASP) \cite{prb.59.1758}. Technically, the generalized gradient approximation (GGA), altogether with the PBE functional for solid (PBE$\_$sol) \cite{prl.100.136406}, were employed since they are known to yield accurate structural parameters \cite{prb.83.094105}. Local density approximation (LDA) was also tested for NaMgF$_3$, yielding  transition pressures merely a few GPa smaller than those from GGA. The projector augmented wave (PAW) method \cite{prb.50.17953} was used to mimic electron-ion interactions with an energy cutoff of 500 eV for all the studied compounds. Hydrostatic pressure was typically simulated up to 120 GPa. 20-atom unit cells were typically adopted for the three investigated structures (i.e., Pv-$Pnma$, pPv-$Cmcm$, ppPv-$Pnma$), with $\Gamma$-centered 6$\times$6$\times$4,  8$\times$4$\times$6 and 6$\times$8$\times$4 $k$-point meshes, respectively. Note, however, that 40-atom cells with a doubled $b$ lattice parameter have also been employed when studying the collinear magnetic properties of the ppPv-$Pnma$ phase of GdFeO$_3$ depicted in Fig. 3 of the manuscript. In that case, the $\Gamma$-centered 6$\times$6$\times$4 $k$-point mesh was used and the $f$ electrons were included in valence to investigate the contribution of Gd to the magnetic properties.  All the studied structures were fully relaxed during the simulations. For instance,  the Hellman-Feynman forces were systematically smaller than 0.001 eV/{\AA} when using 20-atom unit cells, which then allowed to compute elastic constants  in VASP (using an atomic displacement of 0.015 {\AA} and each step being converged to $10^{-8}$ eV).
For all the investigated compounds having the ppPv-$Pnma$ structure (see Fig. 1 of the manuscript), the phonon spectra were calculated using the PHONOPY software \cite{prb.78.134106}. In that case, a 2$\times$3$\times$1 supercell with 120 atoms was employed to compute the force constants, and the $\Gamma$ point was used to sample the Brillouin zone (the validity of such $k$-point sampling was checked by testing  4$\times$4$\times$4 $\Gamma$-centered $k$-point mesh). The energy was converged to $10^{-8}$ eV.
Other technical details include (i) the typical treatment of 4f electrons in the core (except for the results of Fig. 3 of the manuscript) if the $A$ ion of the $AB$O$_3$ material is a rare-earth ion, based on our calculation results showing that freezing such electrons in the core {\it versus} considering them
inside the valence only results in a 0.5\% underestimation of the lattice constants for the ppPv-$Pnma$ phase of GdFeO$_3$;
and (ii) the use of an effective Hubbard U parameter  when ions possess localized $d$ electrons. Typically, we employ $U$ = 4 eV for the $3d$ electrons of Ti, Cr, Mn, Fe and Co \cite{prb.83.094105,prb.77.045137}, and $U$ = 2.8 eV for the $4d$ electrons of Rh, Ru  as well as for the $5d$ electrons of Ir and Pt \cite{prb.88.144102}. Note that we also tested $U$ = 0 eV, 4 eV and 5.3 eV for CaMnO$_3$. These tests show that varying the Hubbard U parameter only makes a difference in the lattice constants of the order of 0.5\% for the ppPv-$Pnma$ phase and has no qualitative influence on the phase diagram (transition pressures were only found to shift by a few GPa).

{\it Crystallographic structure and elastic constants:}

In Table I we report an example of crystallographic structures associated with Pv-$Pnma$, pPv-$Cmcm$ and  ppPv-$Pnma$ phases. Such data can be used  by computational scientists to, e.g., calculate properties of our newly discovered ppPv-$Pnma$ structure or by experimentalists to, e.g., check if X-ray or neutron patterns (they obtained or will obtain) can be well fitted by one of these three phases.

Table II shows a specific hierarchy between elastic constants, such as the facts that C$_{11}$ is much smaller than C$_{22}$ and C$_{33}$ in the Pv-$Pnma$ phase while it is C$_{22}$ that is the smallest (respectively, largest) among the C$_{ii}$ coefficient in pPv-$Cmcm$ (respectively, ppPv-$Pnma$) -- as consistent with their crystallographic structures. Such information can be useful to, e.g., earth-scientists since  it has connection with seismic anisotropy \cite{nature.430.445,nature.438.1142}.

{\it Phonon spectra and electronic structures:}

Figure 1 of this SM shows the phonon spectra and electronic band structures of the ppPv-$Pnma$ phase for NaMgF$_3$ and GdFeO$_3$ under 60 GPa, to demonstrate the stability of this phase as well as the fact that the electronic band gap can have different magnitude and nature (i.e., direct versus indirect) for different materials adopting this ppPv-$Pnma$ phase.

Table III is also reported here to emphasize that the presently discovered ppPv-$Pnma$ structure can have  a wide range of electronic band-gap, implying that the systems adopting it can either be metallic, semiconductor or strong insulator -- which  may broaden their possible range of technological application (especially, if tthe ppPv-$Pnma$ phase can be quenched to ambient conditions).

{\it Phase diagram of  rare-earth ferrites:}

Furthermore, Fig. 2 of the SM displays the phase diagram of the rare-earth ferrites ($R$FeO$_3$) materials revealing the dependence of transition pressure with rare-earth ionic radius.  This is done in order to potentially guide experimentalists in their search of pPv-$Cmcm$ and ppPv-$Pnma$ states in these compounds, as well as to illustrate that their properties do depend on the size of $R$ ion -- as already known for other quantities  \cite{White1969,Hongjian2013,prb.89.205122}.

{\it Comparison between our predictions and previous works:}

{\it 1) For NaMgF$_3$:}

Figure 3 of this SM provides our predicted X-Ray Diffraction (XRD) pattern for the Pv-$Pnma$, pPv-$Cmcm$ and  ppPv-$Pnma$ phases of NaMgF$_3$ at 55 GPa, assuming that the incident X-ray has a wavelength of 0.3344 \AA. This particular wavelength was chosen because it is the one experimentally used in Refs. \cite{martin,martin2}. Our predicted XRD spectra for the pPv-$Cmcm$ phase agrees rather well with the corresponding experimental one (for the post-perovskite structure) shown in the inset of Fig. 3 of Ref. \cite{martin2} for NaMgF$_3$ under a hydrostatic pressure of 55 GPa, which further attests the accuracy of the simulations. Such agreement is further emphasized in Table IV that reports the the calculated and experimental peak positions of the pPv-$Cmcm$ phase at 55 GPa.


Regarding the comparison between our simulations for the XRD pattern of the ppPv-$Pnma$ phase and the  XRD pattern of the N-phase of NaMgF$_3$  measured at 55 GPa after laser heating
 (which is shown in the inset of Fig. 3 in Ref. \cite{martin2}), one first has to know that our predicted XRD of the ppPv-$Pnma$ corresponds to a $\simeq$ 2.3\% increase of the lattice constants with respect to those predicted by our 0 K first-principles calculations. This increase was done in order to have the same volume than in the measurements, and may reflect thermal expansion  (since a {\it high temperature} of $\sim$2000 K was experimentally found necessary for the N-phase to emerge, likely due to kinetic reason) or some inaccuracy in the calculations. Interestingly, such increase
does not qualitatively affect the XRD pattern but was rather found to ``only'' shift the XRD peaks by small amounts. Such small shifts lead to a rather good agreement between the position of many of our simulated peaks and the position of the experimental peaks that were believed to originate from the N-phase in Ref. \cite{martin2} -- as evidenced in Fig. 3(c) of this SM and  in Table IV. Such agreement is  a strong evidence that the experimentally observed N-phase of Refs. \cite{martin,martin2} is our predicted ppPv-$Pnma$ phase. Interestingly, Figure 3 and Table IV of this SM also provide additional information. For instance, it suggests that the rather weak peak at 4.5$^{\circ}$ that was believed to be due to the N-phase in Ref. \cite{martin2} can in fact originate from the pPv-Cmcm state -- since one can see such peak in Fig. 3(b) but not in Fig. 3(c) of this SM. Similarly, Ref. \cite{martin2} assumed that the series of peak they found in the 7$^{\circ}$ - 7.5$^{\circ}$, 9.5$^{\circ}$ - 10$^{\circ}$ and 12.8$^{\circ}$ - 14$^{\circ}$ ranges of angle are caused by the post-perovskite pPv-$Cmcm$ phase, while these peaks also exist in the post-post-perovskite phase (they are the peaks that do not possess corresponding arrows in Fig. 3(c) of this SM).

{\it 2) For NaZnF$_3$:}

 Figure 1(a)  of the manuscript shows that our calculations yield, in the NaZnF$_3$ compound, (i) a transition from  Pv-$Pnma$ to pPv-$Cmcm$ at 8.1 GPa, as rather consistent with the recent measurement of Ref.\cite{jssc.182.1545} yielding the same transition at around 14 GPa; and (ii) a transition from pPv-$Cmcm$ to ppPv-$Pnma$ at 20.3 GPa, which suggests that the unidentified phase that has been observed  to coexist with pPv-$Cmcm$  for pressure above 25 GPa in NaZnF$_3$ \cite{jssc.182.1545}  may be our presently predicted ppPv-$Pnma$. Such hint is strengthened when realizing that Ref.\cite{jssc.182.1545} reported  a large elastic anisotropy  in  NaZnF$_3$ above 25 GPa, which is consistent with our numerical finding that
the elastic (stiffness) constant  in  the ppPv-Pnm phase of any material should be larger along the $b$-axis  than along the $a$ or $c$ axis due to the fact that this $b$-axis is precisely the direction of elongation of the one-dimensional chains  (see Table II of the SM for NaMgF$_3$).
However, the comparison between our predicted XRD spectra of ppPv-$Pnma$ for NaZnF$_3$ shown in Fig. 4(c) of this SM (for a pressure of 30 GPa and an incident wavelength of 0.3738 \AA) and  the experimental XRD pattern depicted in Fig. 2 of Ref. \cite{jssc.182.1545} can not allow us to fully guarantee that the unidentified phase of Ref.  \cite{jssc.182.1545} is our discovered  ppPv-Pnm phase. This is because it is difficult to know what are  the  peaks solely due to this unidentified phase in the pattern shown in Fig. 2 of Ref. \cite{jssc.182.1545} (as a result  of the coexistence between different phases). However, it is promising to realize that our simulated XRD spectra of ppPv-$Pnma$ (see Fig. 4(c) of this SM)  possesses a relatively strong peak for an angle of around 5.5$^{\circ}$, and as observed in Ref. \cite{jssc.182.1545}, such latter peak is inconsistent with the pPv-$Cmcm$ state but is rather associated with this  unidentified phase. We thus hope that the simulated XRD presented in Fig. 4(c) will soon allow the definite experimental identification of this latter phase found in Ref. \cite{jssc.182.1545}.

{\it 3) For MgSiO$_3$:}

Figure 1(b) of the manuscript shows that ppPv-$Pnma$ is not predicted to form  in MgSiO$_3$ when the pressure is applied up to 120 GPa (note that we also conducted further calculations up to 400 GPa in this specific material, and still did not  found this ppPv-$Pnma$ phase). In this important material present in the lower earth mantle, only a  single Pv-$Pnma$--to--pPv-$Cmcm$ transition is found at  95.5 GPa, which is in reasonable agreement with the known experimental value of $\simeq$ 125 GPa happening at high temperature \cite{science.304.855} and with previous first-principles calculations yielding 83.7 GPa or 98.7 GPa   for the critical pressure \cite{nature.430.445} of that precise transition.

{\it 4) For Mn$_2$O$_3$:}

Mn$_2$O$_3$ is also predicted to undergo the single Pv-$Pnma$--to--pPv-$Cmcm$ transition, but at a smaller pressure of 25.3 GPa -- which agrees rather well with the experimental data of Ref.\cite{Santillan} reporting a critical pressure being in the 27-36 GPa range at room temperature.

{\it 5) For Al$_2$O$_3$:}

It is also worthwhile to realize that our calculations yield a Pv-$Pnma$--to--pPv-$Cmcm$ transition in Al$_2$O$_3$ for a pressure of 95.5 GPa, which slightly underestimates the critical pressure of 130 GPa at which the pPv-$Cmcm$ is known to occur  in this compound \cite{epsl.246.326,gpl.32.l16310}.

{\it Discussion about other phases and decomposition}

This latter underestimation between the theoretical and experimental pressures at which pPv-$Cmcm$ forms in Al$_2$O$_3$ is likely due to the fact that, experimentally, the phase adopted by  Al$_2$O$_3$ before the transition is the so-called Rh$_2$O$_3$(II)-type state rather than the Pv-$Pnma$ phase. Such discrepancy between measurements and our calculations takes its origin from the fact that we exclusively concentrate on three phases here (i.e., Pv-$Pnma$, pPv-$Cmcm$ and
ppPv-$Pnma$) for  all our investigated compounds -- in order to demonstrate their possible stability in a wide range of materials. As a result, some transitions may be missed by the simulations or one of the three investigated states can occur at a different pressure with respect to experiments in a few compounds (namely, those for which neither Pv-$Pnma$ nor pPv-$Cmcm$ and  ppPv-$Pnma$ are the most stable state between 0 and 120 GPa). For instance, Fig. 5 of this SM indicates what happens to the phase transitions summarized in Fig. 1 of the manuscript, when  the P2$_1$/m phase discovered in CaRhO$_3$\cite{Shikaro} is also taken into account in the simulations. In that case, out of the thirty six compounds we investigated, eight exhibit some modifications for their phase diagram: P2$_1$/m appears, in some pressure range, in-between the  Pv-$Pnma$ (respectively, pPv-$Cmcm$) and ppPv-$Pnma$ structures for ErFeO$_3$ and TmFeO$_3$ (respectively, LuFeO$_3$, CaTiO$_3$, CaRhO$_3$, CaPtO$_3$ and CaIrO$_3$), while it exists from around 86 GPa up to our highest investigated pressure of 120 GPa in InFeO$_3$.
It is also interesting to notice that comparing Fig. 3 of Ref.  \cite{Shikaro} with Figs. 2(c) and 2(d) of the manuscript suggests that this P2$_1$/m phase can be thought as being a structural bridge between the known pPv-$Cmcm$ state and the presently discovered ppPv-$Pnma$ phase, since it possesses edge-sharing and corner-sharing octahedra sheets (as in pPv-$Cmcm$) but also chains propagating along the $b$-axis (as in ppPv-$Pnma$).

One should also be aware that some previous first-principles calculations \cite{Umemoto1,Umemoto2} suggested that some ABX$_3$ materials, namely MgSiO$_3$ or NaMgF$_3$, may decompose at high pressure into a AX+BX$_2$ mixture, which can therefore render the observation of some of our phases challenging. However, such proposed decomposition was predicted to arise in MgSiO$_3$ at much higher pressure (namely, above 1000 GPa) than the ones presently investigated and was not experimentally found in NaMgF$_3$ up to 70 GPa \cite{Rocholski}. Moreover, it is also important to realize that states that are {\it not} ground state can be created and then stabilized in many materials by various techniques {\it if these states are metastable and that a large kinetic barrier separates them with the ground state} -- as it is the case for ppPv-$Pnma$ (see our discussion about dynamic stability in the manuscript).
We are therefore confident that our present results will be soon experimentally confirmed.

\clearpage
\begin{table}[h]\centering  
\caption{Crystallographic structures of NaMgF$_3$ in the three investigated phases (Pv-$Pnma$,
pPv-$Cmcm$ and  ppPv-$Pnma$), at 60 GPa. Mg-F bond lengths within the octahedra and their degeneracies (indicated in parenthesis) are also given for information.}
\begin{tabular}{cccccc}
\hline\hline
\multirow{2}{*}{Pv-$Pnma$} & &a = 5.167 \AA &b = 6.805 \AA &c = 4.299 \AA &\\
 & &\multicolumn{3}{c}{$\alpha$ = $\beta$ = $\gamma$ = 90$^\circ$} &\\
\hline
Atom & Wyc. & x & y & z & Mg-F bond length\\
F  & 8d & 0.3193 & 0.5644 & 0.6477 & \\
F  & 4c & 0.5412 & 0.25   & 0.8673 & 1.807($\times$2)\\
Mg & 4b & 0      & 0      & 0.5    & 1.832($\times$2)\\
Na & 4c & 0.8939 & 0.25   & 0.0510 & 1.821($\times$2)\\
\hline
\hline
\multirow{2}{*}{pPv-$Cmcm$} & &a = 2.713 \AA &b = 8.154 \AA &c = 6.766 \AA &\\
 & &\multicolumn{3}{c}{$\alpha$ = $\beta$ = $\gamma$ = 90$^\circ$} &\\
\hline
Atom & Wyc. & x & y & z & Mg-F bond length\\
F  & 8f & 0 & 0.64457  & 0.43765 &  \\
F  & 4c & 0 & -0.06993 & 0.25    &  1.785($\times$2)\\
Mg & 4a & 0 & 0        & 0       &  1.846($\times$4)\\
Na & 4c & 0 & 0.25039  & 0.25    &  \\
\hline
\hline
\multirow{2}{*}{ppPv-$Pnma$} & &a = 5.082 \AA &b = 2.829 \AA &c = 10.215 \AA &\\
 & &\multicolumn{3}{c}{$\alpha$ = $\beta$ = $\gamma$ = 90$^\circ$} &\\
\hline
Atom & Wyc. & x & y & z & Mg-F bond length\\
F  & 4c & 0.0036 & 0.25 & 0.8980 & \\
F  & 4c & 0.1573 & 0.25 & 0.2823 & 1.813\\
F  & 4c & 0.8600 & 0.25 & 0.4504 & 1.942\\
Mg & 4c & 0.7739 & 0.25 & 0.0499 & 1.871($\times$2)\\
Na & 4c & 0.6266 & 0.25 & 0.8007 & 1.888($\times$2)\\
\hline\hline
\end{tabular}
\label{tab: struc1}
\end{table}

\clearpage

\begin{table}[h]
\centering
\caption{Elastic constants of NaMgF$_3$  for the three investigated Pv-$Pnma$,
pPv-$Cmcm$ and  ppPv-$Pnma$ phases, for a hydrostatic pressure of 0 GPa and 60 GPa. These elastic constants are given in GPa units and the `1', `2' and `3' subscripts refer to the a-, b- and c-axis, respectively, of each of these three phases.}
\begin{tabular}{ccccccc}
\hline\hline
$\qquad$$\qquad$&\multicolumn{2}{c}{$\quad$ Pv-$Pnma$ $\qquad$}&\multicolumn{2}{c}{$\quad$ pPv-$Cmcm$ $\qquad$}&\multicolumn{2}{c}{$\quad$ ppPv-$Pnma$ $\qquad$}\\\cline{2-7}
&0 GPa &60 GPa &0 GPa &60 GPa &0 GPa &60 GPa\\
\hline
$C_{11}$ &137   &224 	&179 &431 	&65 	&324\\
$C_{22}$ &159   &477 	&111 &271 	&163 	&401\\
$C_{33}$ &155   &386 	&170 &458 	&117 	&358\\
$C_{44}$ &49 	  &126 	&28  &104	  &25 	&111\\
$C_{55}$ &43 	  &81 	&28  &102	  &36 	&130\\
$C_{66}$ &53 	  &114 	&38  &198 	&39 	&210\\
$C_{12}$ &58 	  &220 	&34  &226 	&33 	&233\\
$C_{13}$ &42 	  &197 	&52  &151 	&46 	&185\\
$C_{23}$ &46 	  &169 	&27  &232 	&28 	&207\\
\hline\hline
\end{tabular}
\label{tab: struc1}
\end{table}

\clearpage
\begin{table}[h]
\centering
\caption{Electronic band gaps of the investigated materials in their ppPv-$Pnma$ phase, just above their transition pressure. The nature of the band gap is indicated inside the parenthesis in the second column. Note that we numerically found (not shown here) that our predicted band gap of  BiFeO$_3$ in its $R3c$ ground state  is 2.1 eV, while the corresponding experimental  value is close to 2.5 eV \cite{apl.89.102506}. Such comparison indicates that, as common with  first-principles-calculations, our predicted band gaps in this Table III likely underestimate measurements by around 0.4 eV. Note that we also performed hybrid functional calculations \cite{HSE} (not shown here) and found that, for BFO systems, they overestimate the band gap by $\simeq$0.8eV.}
\scriptsize
\begin{tabular}{cccc}
\hline\hline
\multirow{2}{*}{Compounds}& \multirow{2}{*}{ $\qquad$Gap (in eV)$\qquad$ } &  Pressure (in GPa)   &  Pressure (in GPa)   \\
&&at which the gap is calculated&at which  ppPv-$Pnma$ appears\\
\hline
NaCoF$_3$    &    0.21 (direct)       &    20   &  14.8     \\
NaZnF$_3$    &    4.64 (direct)       &    30   &  20.3     \\
NaNiF$_3$    &    3.56 (direct)       &    80   &  78.7     \\
NaMgF$_3$    &    9.04 (direct)       &    60   &  51.1     \\
LuFeO$_3$    &    0.25 (indirect)       &    70   &  64.9     \\
InFeO$_3$    &    $\sim$0 (indirect)    &    120  &  118      \\
CaTiO$_3$    &    1.89 (direct)       &    80   &  70.8     \\
CaCrO$_3$    &    0.42 (direct)       &    60   &  60       \\
MnGeO$_3$    &    metallic   &    110  &  110      \\
Fe$_2$O$_3$  &    0.46 (indirect)       &    90   &  80.7     \\
CaMnO$_3$    &    1.25 (indirect)       &    30   &  23.6     \\
TmFeO$_3$    &    0.36 (indirect)       &    60   &  54.6     \\
ErFeO$_3$    &    0.39 (indirect)       &    60   &  53.8     \\
HoFeO$_3$    &    0.38 (indirect)       &    60   &  53.6     \\
DyFeO$_3$    &    0.44 (indirect)       &    60   &  53.6     \\
TbFeO$_3$    &    0.44 (indirect)       &    60   &  54.4     \\
GdFeO$_3$    &    0.83 (indirect)       &    60   &  56.5     \\
SmFeO$_3$    &    0.51 (indirect)       &    80   &  71.8     \\
PmFeO$_3$    &    0.50 (indirect)       &    100  &  90.7     \\
BiCrO$_3$    &    $\sim$0 (indirect)    &    60   &  50.5     \\
BiFeO$_3$    &    $\sim$0 (indirect)    &    100  &  98.2     \\
CaPtO$_3$    &    1.56 (indirect)       &    90   &  81       \\
CaIrO$_3$    &    metallic   &    90   &  86.5     \\
CaRuO$_3$    &    0.54 (direct)       &    40   &  33.8     \\
CaRhO$_3$    &    metallic   &    70   &  64.5     \\
\hline\hline
\end{tabular}
\label{tab: sbandgap}
\end{table}

\clearpage

\begin{table}[h]
  \centering
  \caption{Comparison between the measurements of Ref. \cite{martin2} and our present simulations for some XRD  peak positions (in degrees) of the pPv-$Cmcm$ state and of another phase of NaMgF$_3$ at 55 GPa, for an incident X-ray having a wavelength $\lambda$ = 0.3344 \AA. This other phase is ppPv-$Pnma$ in the simulations while it is denoted as the N-phase in the experiments of Refs. \cite{martin,martin2}.    Note that the  lattice constants are expanded by 2.3\% with respect to their predicted 0 K values for the computation of the theoretical  XRD pattern of ppPv-$Pnma$ to reproduce the experimental volume of the N-phase. Note also that the experimental peak positions of the N-phase  reported in this Table are those {\it believed} to originate from this N-phase in Ref. \cite{martin2}, but that other peaks have also been measured in the XRD spectra of Ref. \cite{martin2} (see our discussion in the SM). The peaks are indexed following the order of increasing 2$\theta$ (in Deg.), as indicated in the inset of Fig. 3 of Ref. \cite{martin2}.}
  \scriptsize
  \begin{tabular}{cccccc}
    \hline\hline
    \multirow{2}{*}{Peak index} &\multicolumn{2}{c}{pPv-$Cmcm$} & &\multicolumn{2}{c}{N-phase/ppPv-$Pnma$} \\\cline{2-3}\cline{5-6}
    & Experiment \cite{martin2}  & Theory && Experiment \cite{martin2}  & Theory\\
    \hline
    1 &-     & 4.6& &4.5   &  -- \\
    2 &5.6   & 5.6& &5.0   &  5.1\\
    3 &7.2   & 7.3& &6.9   &  6.8\\
    4 &7.4   & 7.4& &7.9   &  8.1\\
    5 &7.9   & 7.9& &8.0   &  8.3\\
    6 &9.6   & 9.7& &8.2   &  8.5\\
    7 &9.8   & 9.9& &8.8   &  9.1\\
    8 &10.2  &10.3& &9.0   &  9.3\\
    9 &11.2  &11.3& &10.2  &  10.4\\
    10& - & -& &10.7   &  10.9\\
    11& - & -& &11.1   &  11.2\\
    12& - & -& &11.4   &  11.6\\
    13& - & -& &11.6   &  11.8\\
    14& - & -& &12.0   &  12.2\\
    15& - & -& &12.6   &  12.9\\
    16& - & -& &14.4   &  14.7\\
    \hline\hline
  \end{tabular}
  \label{tab:xrd}
\end{table}

\clearpage

\begin{figure}  [h] 
\includegraphics[width=15cm]{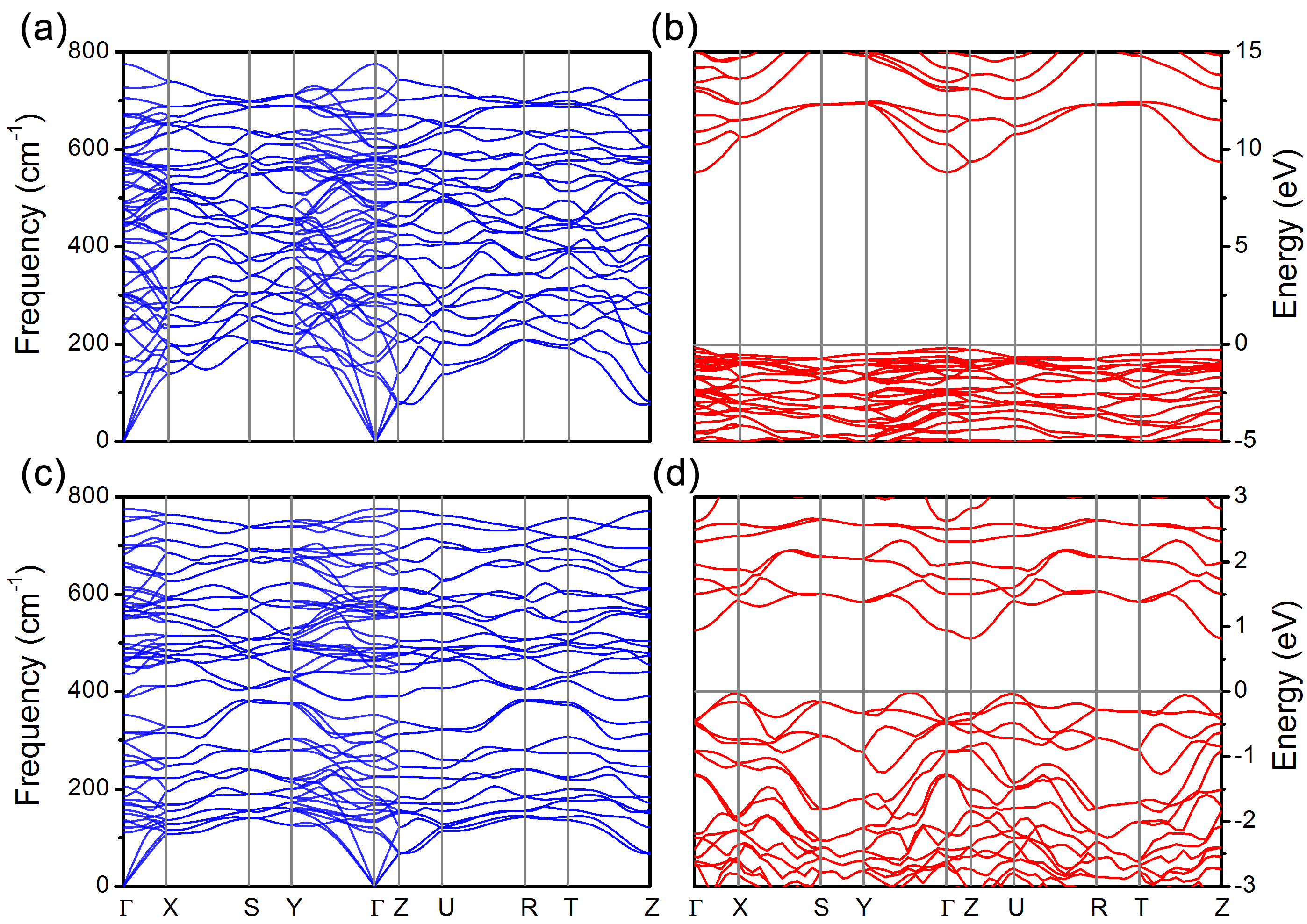}
\caption{(Color online). Phonon spectrum and electronic band structure of the ppPv-$Pnma$ phase for NaMgF$_3$ (Panels a and b) and GdFeO$_3$ (Panels c and d) under 60 GPa. Note that all the phonons have positive square frequencies (i.e., they are stable) in both NaMgF$_3$ and GdFeO$_3$. Note also that the electronic band gap is direct  at the zone-center in NaMgF$_3$ while it is indirect (from $\frac{2\pi}{b}$(0 $\frac{1}{4}$ 0) to $\frac{2\pi}{c}$(0 0 $\frac{1}{2}$)) in GdFeO$_3$.}
\end{figure}

\begin{figure}
\includegraphics[width=12cm]{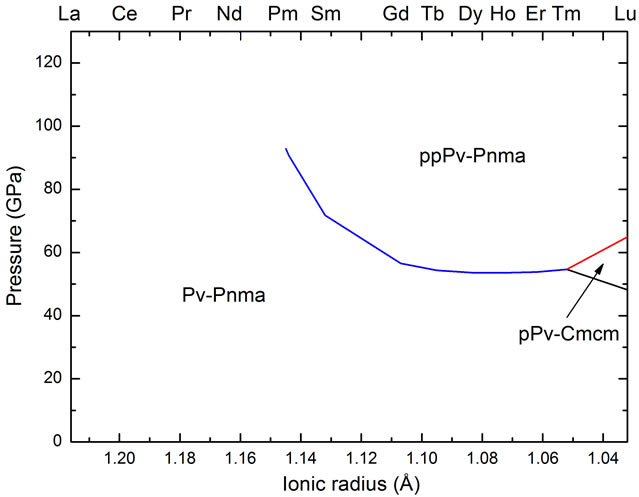}%
\caption{(Color online). Phase diagram of the $R$FeO$_3$ rare-earth ferrites representing critical pressures as a function of the rare-earth ionic radius. This phase diagram only concerns the Pv-$Pnma$, pPv-$Cmcm$ and  ppPv-$Pnma$ phases, up to 120 GPa.}
\end{figure}

\begin{figure}[h]
\includegraphics[width=12cm]{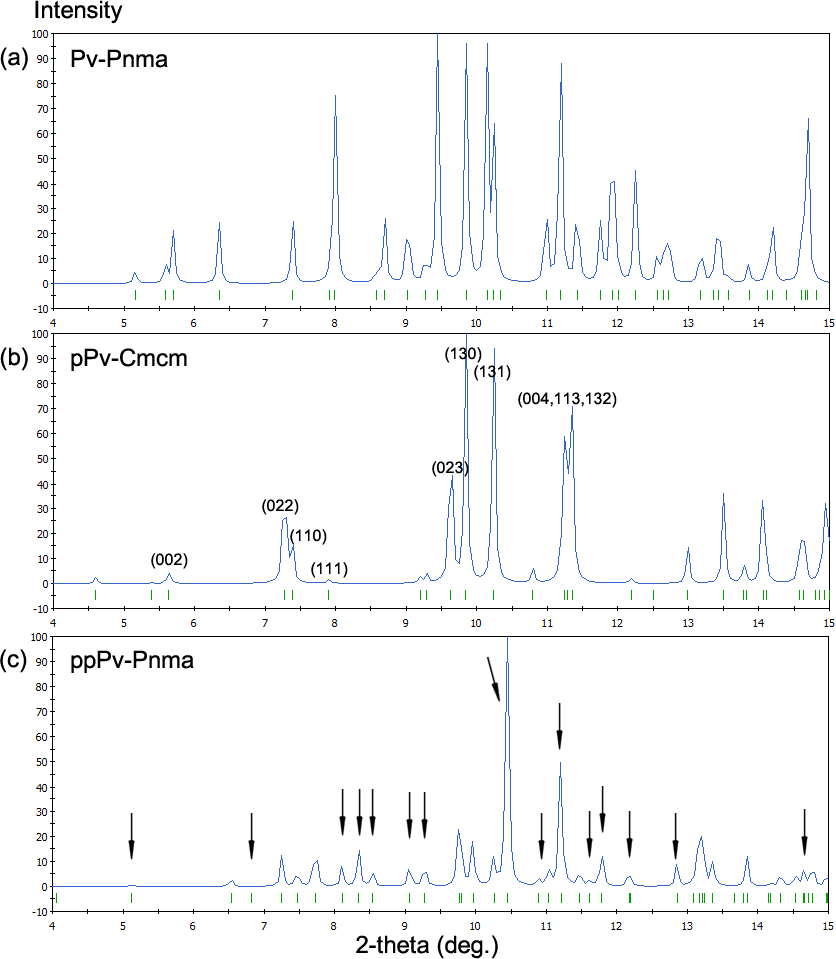}
\caption{(Color online). Simulated X-Ray Diffraction pattern for polymorphs Pv-$Pnma$ (Panel a), pPv-$Cmcm$ (Panel b), and ppPv-$Pnma$ (Panel c) of NaMgF$_3$ at 55 GPa, for an incident X-ray having a wavelength $\lambda$ = 0.3344 \AA. The arrows in Panel (c) correspond to the position of the peaks that have been measured and tentatively assigned to the N-phase of NaMgF$_3$ in Ref. \cite{martin2}.}\label{fig:xrd}
\end{figure}

\begin{figure}[h]
\includegraphics[width=12cm]{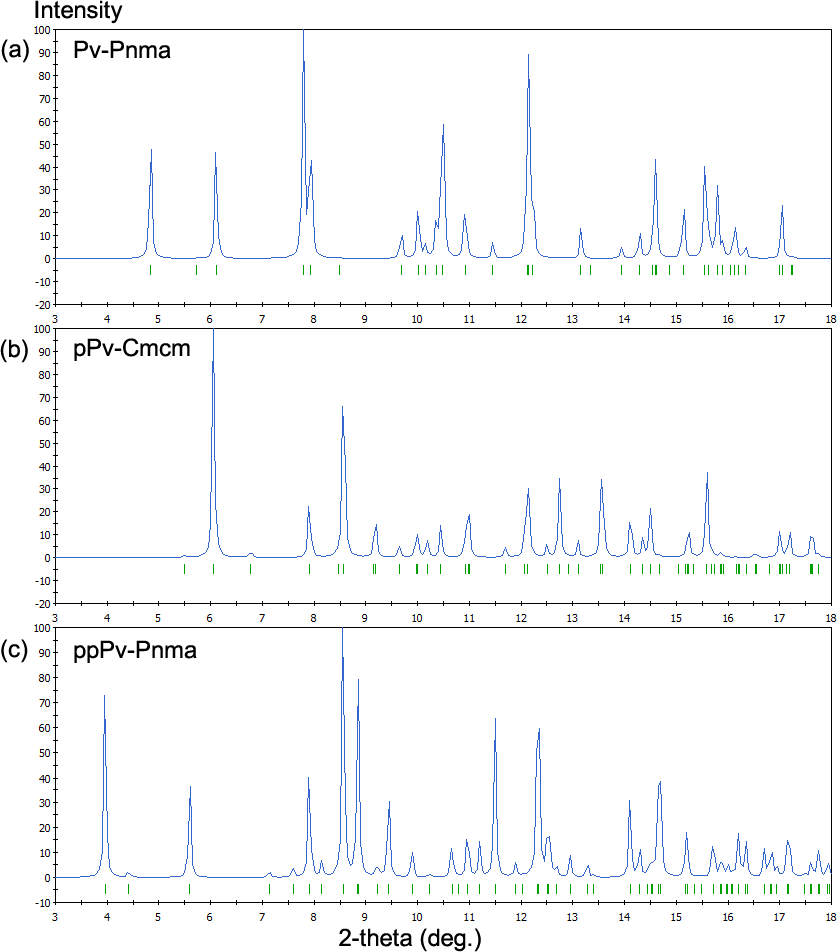}
\caption{(Color online). Simulated X-Ray Diffraction pattern for polymorphs Pv-$Pnma$ (Panel a), pPv-$Cmcm$ (Panel b), ppPv-$Pnma$ (Panel c) of NaZnF$_3$ at 30 GPa, for an incident X-ray having a wavelength $\lambda$ = 0.3738 \AA.}\label{fig:xrd}
\end{figure}

\begin{figure}  [h]
\includegraphics[width=12cm]{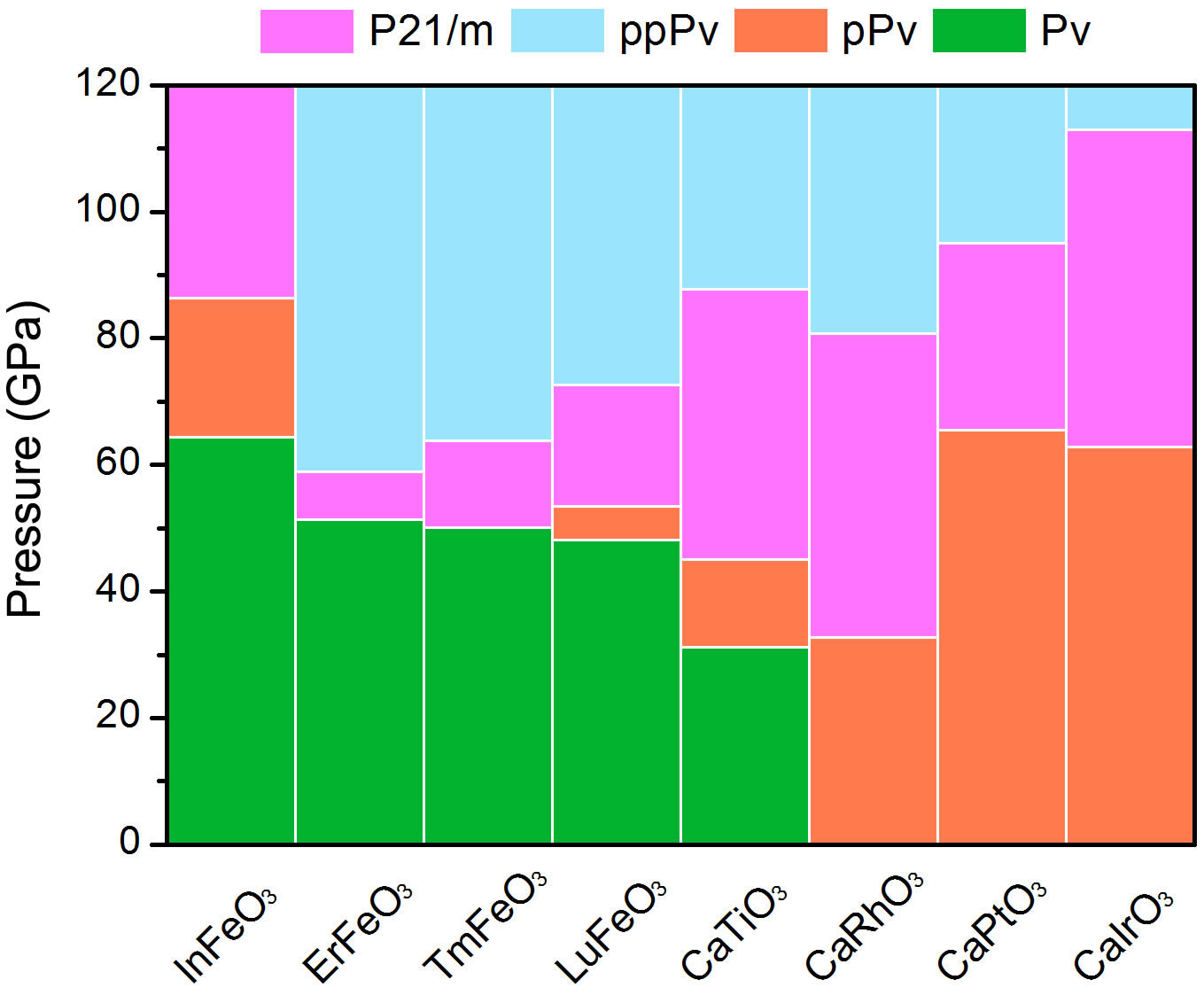}%
\caption{(Color online). Modification of the pressure range of stability of the Pv-$Pnma$, pPv-$Cmcm$ and ppPv-$Pnma$ phases in the {\it ABX$_3$} and $A_2$O$_3$  materials under study, when the P2$_1$/m phase discovered in CaRhO$_3$ \cite{Shikaro} is incorporated into the calculations.}
\end{figure}